\documentclass[english,prl,twocolumn,superscriptaddress,letterpaper]{revtex4}

\usepackage{ae} 
\usepackage[T1]{fontenc}
\usepackage[ansinew]{inputenc}
\usepackage{amsmath}
\usepackage{amssymb}
\usepackage{dsfont}
\usepackage{graphicx}
\usepackage{color}
\usepackage{epsfig}
\usepackage{booktabs}

\def\d{{\rm d}}

\usepackage{varioref}
\usepackage{makeidx}
\makeindex

\usepackage[english]{babel}

\begin{document}

\title{Integration-by-parts reductions from unitarity cuts and algebraic geometry}

\author{Kasper J. Larsen}
\affiliation{Institute for Theoretical Physics, ETH Z{\"u}rich, 8093 Z{\"u}rich, Switzerland}
\author{Yang Zhang}
\affiliation{Institute for Theoretical Physics, ETH Z{\"u}rich, 8093 Z{\"u}rich, Switzerland}

\begin{abstract}
We show that the integration-by-parts reductions of various two-loop integral topologies
can be efficiently obtained by applying unitarity cuts to a specific set of subgraphs
and solving associated polynomial (syzygy) equations.
\end{abstract}

\maketitle

\section{Introduction}

Precise predictions of the production cross sections
at the Large Hadron Collider (LHC) are necessary to gain a
quantitative understanding of the Standard Model signals and background.
To match the experimental precision and the parton distribution
function uncertainties, this typically requires computations
at next-to-next-to leading order (NNLO) in fixed-order perturbation
theory. Calculations at this order are challenging
due to a large number of contributing Feynman diagrams,
involving loop integrals with high powers of loop momenta in the
numerator of the integrand.

A key tool in these computations are integration-by-parts (IBP)
identities \cite{Tkachov:1981wb,Chetyrkin:1981qh}, arising from
the vanishing integration of total derivatives. Schematically,
the relations take the form,
\begin{equation}
\int \prod_{i=1}^L \frac{\d^D \ell_i}{\pi^{D/2}}
\sum_{j=1}^L \frac{\partial}{\partial \ell_j^\mu}
\frac{v_j^\mu \hspace{0.5mm} P}{D_1^{a_1} \cdots D_k^{a_k}}
\hspace{1mm}=\hspace{1mm} 0 \,, \label{eq:IBP_schematic}
\end{equation}
where $P$ and the vectors $v_j^\mu$ are polynomials in the internal and
external momenta, the $D_k$ denote inverse propagators, and
$a_i \geq 1$ are integers. In practice, the IBP identities
generate a large set of linear relations between loop integrals,
allowing most of them to be reexpressed in terms of a set of
so-called master integrals. (The fact that the linear basis of
integrals is always finite was proven in Ref.~\cite{Smirnov:2010hn}.)
The latter step of solving the linear systems arising from
Eq.~(\ref{eq:IBP_schematic}) may be done by Gaussian elimination
in the form of the Laporta algorithm~\cite{Laporta:2000dc,Laporta:2001dd},
leading in general to relations involving integrals with
squared propagators. There are several publically available implementations
of automated IBP reduction: AIR~\cite{Anastasiou:2004vj},
FIRE~\cite{Smirnov:2008iw,Smirnov:2014hma},
Reduze~\cite{Studerus:2009ye,vonManteuffel:2012np},
LiteRed~\cite{Lee:2012cn}, along with private implementations.
A formalism for deriving IBP reductions without squared
propagators was developed in Ref.~\cite{Gluza:2010ws}.
A recent approach \cite{vonManteuffel:2014ixa} uses numerical
sampling of finite-field elements to construct the reduction
coefficients.

In addition to reducing the contributing Feynman diagrams to
a minimal set of basis integrals, the IBP reductions
provide a way to compute these integrals themselves via
differential equations%
~\cite{Kotikov:1990kg,Kotikov:1991pm,Bern:1993kr,Remiddi:1997ny,Gehrmann:1999as,Henn:2013pwa,Ablinger:2015tua}.
Letting $x_m$ denote a kinematical variable,
$\epsilon = \frac{4-D}{2}$ the dimensional regulator, and
$\boldsymbol{\mathcal{I}}(\boldsymbol{x},\epsilon)
=\big(\mathcal{I}_1 (\boldsymbol{x},\epsilon),
\ldots, \mathcal{I}_N  (\boldsymbol{x},\epsilon) \big)$
the basis of integrals, the result of differentiating a basis integral
wrt. $x_m$ can again be written as a linear combination
of the master integrals (in practice, using the IBP reductions).
As a result, one has a linear system of differential equations,
\begin{equation}
\frac{\partial}{\partial x_m} \boldsymbol{\mathcal{I}}(\boldsymbol{x},\epsilon)
= A_m (\boldsymbol{x}, \epsilon) \boldsymbol{\mathcal{I}}(\boldsymbol{x},\epsilon) \,,
\end{equation}
which, supplied with appropriate boundary conditions, can be
solved to yield expressions for the master integrals.
This has proven to be a powerful tool for compu\-ting two-
and higher-loop integrals. IBP reductions thus play a central role
in perturbative calculations in particle physics.

In many realistic multi-scale problems, such as $2 \to n$ scattering
amplitudes with $n\geq 2$, the generation of IBP reductions with
existing algorithms is the most challenging part of the calculation.
It is therefore interesting to explore other methods for generating
these reductions.

In this paper we show that the IBP reductions of various
two-loop integral topologies can be obtained efficiently by applying
unitarity cuts to a specific set of subgraphs
and solving associated polynomial (syzygy) equations.
A similar approach was introduced by Harald Ita in ref.~\cite{Ita:2015tya}
where IBP relations are also studied in connection with cuts, and
their underlying geometric interpretation is clarified.

\section{Method}

Our starting point is a two-loop integral with $n$ external legs
and $k$ propagators. (Note that, after integrand reduction, $k \leq 11$.)
We work in dimensional regularization and use the
FDH scheme, taking the external momenta in four dimensions.
Accordingly, we decompose the loop momenta into four- and
$(D-4)$-dimensional parts, $\ell_i = \overline{\ell}_i + \ell_i^\perp$,
$i=1,2$. Defining $\mu_{ii}\equiv-( \ell_i^\perp)^2 \geq 0$
and $\mu_{12}\equiv -\ell_1^\perp \cdot \ell_2^\perp$, the
two-loop integral then takes the form
\begin{gather}
I^{(2)}_{n\geq 5} = \frac{2^{D-6}}{\pi^{5}\Gamma(D-5) }\int_0^\infty \d \mu_{11} \int_0^\infty \d \mu_{22}
  \int_{-\sqrt{\mu_{11}\mu_{22}}}^{\sqrt{\mu_{11}\mu_{22}}} \d\mu_{12}
 \nonumber \\
 \times \big(\mu_{11} \mu_{22}-\mu_{12}^2 \big)^{\frac{D-7}{2}}  \int \d^4
 \overline{\ell}_1  \hspace{0.5mm}  \d^4 \overline{\ell}_2 \frac{ P(\ell_1,\ell_2)}{D_1
   \cdots D_k} \,.
 \label{two-loop-integral}
\end{gather}
We now introduce a parametrization which will turn out
to be very useful for the study of IBP reductions.
For $n \geq 5$ there are $11-k$ irreducible scalar products (ISPs) which
we denote by $g_j$, $j=1, \ldots, 11-k$. We can then define
variables $z_1,\ldots, z_{11}$ as follows,
\begin{equation}
z_i \equiv \left\{
\begin{array}{lrl}
  D_i     \hspace{2mm} & 1  \leq &\hspace{-1mm} i \leq k\\[1mm]
  g_{i-k} \hspace{2mm} & k+1\leq &\hspace{-1mm} i \leq m \,,
\end{array} \right.
\label{eq:2}
\end{equation}
with $m=11$ for $n\geq5$. It can be shown that, for an arbitrary two-loop diagram,
the transformation $\{\bar \ell_i, \mu_{ij}\} \rightarrow \{z_1, \ldots, z_{11}\}$
is invertible, with a polynomial inverse (provided the
$g_j$ are chosen to take the form
$\frac{1}{2} (\ell_i + K_j)^2$, rather than linear
dot products of the $\ell_i$), and has a constant Jacobian.
The integral (\ref{two-loop-integral}) can then be written as,
\begin{equation}
I^{(2)}_{n\geq 5} = \frac{2^{D-6}}{\pi^{5}\Gamma(D-5) J} \int \prod_{i=1}^{11} \d z_i \hspace{0.6mm}
  F(z)^{\frac{D-7}{2}} \frac{P(z)}{z_1 \cdots z_k}\,,
\label{two-loop-integral-z}
\end{equation}
where $F(z)$ denotes the kernel $(\mu_{11} \mu_{22}-\mu_{12}^2)$
expressed in the new variables, in which it is polynomial. Note that all
denominator factors in Eq.~(\ref{two-loop-integral-z}) are linear in the $z_i$.

For $n\leq 4$ external legs, the loop momenta have components
which can be integrated out before the transformation
(\ref{eq:2}) is applied. For example, when $n=4$, we can define a four-dimensional
vector $\omega$ such that $p_i \cdot \omega=0$, $i=1,\ldots, 4$.
The loop momenta can then be decomposed in projections orthogonal and
parallel to $\omega$: $\overline{\ell}_i = \overline{\ell}_i^{[3]} + \alpha_i \omega$.
Defining
\begin{gather}
  \lambda_{ij}\equiv\mu_{ij}+\frac{(\ell_i\cdot \omega)
  (\ell_j\cdot \omega)}{\omega^2},\quad i,j= 1,2 \,,
\label{lambda}
\end{gather}
integrating out the components $\ell_i \cdot \omega$
in Eq.~(\ref{two-loop-integral}) yields,
\begin{align}
I^{(2)}_4 &= \frac{2^{D-5}}{\pi^{4}\Gamma(D-4) }\int_0^\infty \d \lambda_{11} \int_0^\infty \d \lambda_{22}
  \int_{-\sqrt{\lambda_{11}\lambda_{22}}}^{\sqrt{\lambda_{11}\lambda_{22}}} \d \lambda_{12}
 \nonumber \\
 &\hspace{2mm} \times (\lambda_{11} \lambda_{22}-\lambda_{12}^2)^{\frac{D-6}{2}}  \int  \d^3
 \overline{\ell}_1^{[3]} \d^3 \overline{\ell}_2^{[3]} \frac{ P(\ell_1,\ell_2)}{D_1
   \cdots D_k} \,.
 \label{two-loop-integral-4}
\end{align}
In this case, there are $9-k$ irreducible scalar products, which we label
as $g_i$, $i=1, \ldots, 9-k$. We define new variables
as in Eq.~(\ref{eq:2}), with $m=9$. Again, the map
$\{\bar \ell_i^{[3]}, \mu_{ij}\} \to \{z_1, \ldots, z_9 \}$ has a
constant Jacobian $J$ and a polynomial inverse.
Hence, the integral in Eq.~(\ref{two-loop-integral-4})
can be rewritten in the simple form,
\begin{equation}
I^{(2)}_4 = \frac{2^{D-5}}{\pi^{4}\Gamma(D-4) J} \int \prod_{i=1}^{9} \d z_i \hspace{0.6mm}
  F(z)^{\frac{D-6}{2}} \frac{P(z)}{z_1 \cdots z_{k}} \,.
\label{two-loop-integral-4z}
\end{equation}
For the following discussion, we may drop the prefactors
in front of the integral signs in Eqs.~(\ref{two-loop-integral-z})
and (\ref{two-loop-integral-4z}).

The virtue of the representations (\ref{two-loop-integral-z}) and
(\ref{two-loop-integral-4z}) is that they make manifest
the effect of cutting propagators $D_i^{-1} \to \delta(D_i)$,
a tool we will employ shortly in our study of IBP reductions.
For a given $c$-fold cut ($0\leq c\leq k$), let
$\mathcal{S}_\mathrm{cut}$, $\mathcal{S}_\mathrm{uncut}$ and
$\mathcal{S}_\mathrm{ISP}$ denote the sets of indices labelling
cut propagators, uncut propagators and ISPs, respectively.
(Cf.~the labelling of propagators, e.g., in Fig.~\ref{fig:dbox_cut}.)
$\mathcal{S}_\mathrm{cut}$ thus contains $c$ elements. Moreover, we let $m$
denote the total number of $z_j$ variables, and
set $\mathcal{S}_\mathrm{uncut}=\{r_1,\ldots, r_{k-c}\}$
and $\mathcal{S}_\mathrm{ISP}=\{r_{k-c+1},\ldots, r_{m-c}\}$.
Then, by cutting the propagators, $z_i^{-1} \to \delta(z_i),
i \in \mathcal{S}_\mathrm{cut}$, the integrals (\ref{two-loop-integral-z}) and
(\ref{two-loop-integral-4z}) reduce to,
\begin{equation}
I^{(2)}_\mathrm{cut} = \int \frac{\d z_{r_1} \cdots \d z_{r_{m-c}} P(z)}
{z_{r_1}\cdots z_{r_{k-c}}} F(z)^{\frac{D - h}{2}} \bigg|_{z_i=0\,, \forall i\in \mathcal{S}_\mathrm{cut}} \,,
\label{cut-z}
\end{equation}
where $h$ is a constant which depends on the number
of external legs: $h=6$ for $n=4$ and $h=7$ for
$n \geq 5$.


We now turn to IBP relations. An IBP relation (\ref{eq:IBP_schematic})
concerning integrals with $m$ integration variables
corresponds to a total derivative or, equivalently,
an exact differential form of degree $m$. We can
likewise consider a $c$-fold cut of an IBP relation,
in which the propagators of $\mathcal{S}_\mathrm{cut}$ are
put on shell in all terms (and integrals which do not contain all these propagators
are set to zero). Such $(m-c)$-fold cut IBP relations
correspond to exact differential forms of degree $m-c$.
In both cases, we can find such differential forms
in a systematic way. For example, the generic exact form, which matches
the form of the integrand in Eq.~(\ref{cut-z}), is,
\begin{gather}
  \label{ansatz}
0\hspace{-0.5mm}=\hspace{-0.5mm}\int \hspace{-0.8mm}
\d \bigg( \hspace{-0.4mm} \sum_{i=1}^{m-c} \hspace{-0.8mm}
     \frac{(-1)^{i+1} a_{r_i} F(z)^{\frac{D-h}{2}}}{z_{r_1}\cdots z_{r_{k-c}}}
     \d z_{r_1} \hspace{-0.5mm} \wedge \hspace{-0.5mm} \cdots \widehat{\d z_{r_i}} \cdots
     \hspace{-0.5mm} \wedge \hspace{-0.5mm} \d z_{r_{m-c}} \hspace{-0.5mm} \bigg)
\end{gather}
{\vskip -1.5mm}
\noindent where the $a_i$'s are polynomials in $\{z_{r_1}, \ldots,
z_{r_{m-c}}\}$. (Similar differential form ans{\"a}tze for four-dimensional
IBPs on cuts were considered in ref.~\cite{Zhang:2014xwa}.) Expanding Eq.~(\ref{ansatz}),
we get an IBP relation,
\begin{gather}
  0=\int \bigg(\sum_{i=1}^{m-c} \Big(\frac{\partial a_{r_i} }{\partial z_{r_i}}
  + \frac{D-h}{2F}a_{r_i}\frac{\partial F }{\partial
    z_{r_i}}\Big)-\sum_{i=1}^{k-c} \frac{a_{r_i}}{z_{r_i}}\bigg) \nonumber \\
\hspace{7mm} \times \frac{F(z)^{\frac{D-h}{2}}}{z_{r_1}\cdots z_{r_{k-c}}} \d z_{r_1} \wedge \cdots
  \wedge \d z_{r_{m-c}} \,. \label{IBP}
\end{gather}
Note that, in general, the second term in the sum gives an integral in
$(D-2)$ dimensions, while the third term generates integrals with doubled
propagators. To get an IBP relation in $D$ dimensions with single-power
propagators, we require that the poles $1/F$ and $1/z_{r_i}$ in the sum cancel,
\begin{align}
b F + \sum_{i=1}^{m-c} a_{r_i}\frac{\partial F}{\partial
z_{r_i}} &=0 \label{eq:syzygy_1} \\
a_{r_i} + b_{r_i} z_{r_i}&=0\,, \hspace{4mm} i=1,\ldots, k-c \,,
\label{eq:syzygy_2}
\end{align}
where $a_{r_i}$, $b$ and $b_{r_i}$ must be polynomials in $z_j$.
Equations of this kind are known in algebraic geometry as
\emph{syzygy equations}. They were previously considered in the
context of IBP relations in Refs.~\cite{Gluza:2010ws,Schabinger:2011dz}.
However, whereas the computations of those references
involve syzygies of \emph{module elements}
(i.e., systems of polynomial equations), here only syzygies of \emph{polynomials}
are required. As the latter are much more efficient to compute,
our method simplifies the generation of IBPs. To see that only
syzygies of polynomials are involved here, note that the last $(k-c)$
equations in Eq.~(\ref{eq:syzygy_2}) are trivial since they are solved by
$a_{r_i}=-b_{r_i} z_{r_i}$. We therefore only have one syzygy equation
to solve,
\begin{equation}
   b F -\sum_{i=1}^{k-c} b_{r_i}z_{r_i}\frac{\partial F }{\partial z_{r_i}} \hspace{0.8mm}+\hspace{0.8mm}
    \sum_{j=k-c+1}^{m-c} a_{r_j}\frac{\partial F}{\partial z_{r_j}}
    \hspace{0.8mm}=\hspace{0.8mm} 0
\label{syz}
\end{equation}
for $m-c+1$ polynomials $b_{r_i}$, $a_{r_i}$ and $b$.
We solve Eq.~(\ref{syz}) by the computational algebraic
geometry software Macaulay2~\cite{M2} or Singular~\cite{DGPS}.
The corresponding IBP relation is then obtained by
plugging the solutions of the syzygy equations into
Eq.~(\ref{IBP}).

In our approach, we can generate IBP identities either directly,
without applying cuts, or generate them on various cuts and then merge the results to
complete IBPs.

In practice, we start by applying our method numerically, with
rational or finite-field coefficients, to determine the set of master
integrals. (Similar ideas for finding a basis of integrals have
appeared in Ref.~\cite{RobRadCor2013}, using random prime numbers for the
external invariants and spacetime dimension, and in Ref.~\cite{Kant:2013vta},
using finite-field elements.)
Then we proceed analytically and find the IBP reductions
on a specific set of cuts, the latter dictated by the set of master
integrals, as explained in the next section. Finally, we
merge the cut results to get complete IBP reductions.

Besides the systematic approach using syzygies, we also remark that
simple ans{\"a}tze such as
\begin{align}
  \label{eq:9}
  0&=\int \d\bigg(\omega_{m-k-2} \wedge \d F\wedge \d z_1 \cdots \wedge
  \d z_k\frac{F^\frac{D-h}{2}}{z_1 \cdots z_k}\bigg) \,, \nonumber \\
0&=\int \d \bigg(\omega_{m-k-1} \wedge \d z_1 \cdots\wedge
  \d z_k\frac{F^\frac{D-h+2}{2}}{z_1 \cdots z_k}\bigg) \,,
\end{align}
easily generate a large portion (but not all) of the IBP relations
without doubled propagators in $D$ dimensions. Here $\omega_l$
stands for an arbitrary polynomial-valued $l$-form.

\section{Example}

\begin{figure}[!h]
\includegraphics[scale=0.8]{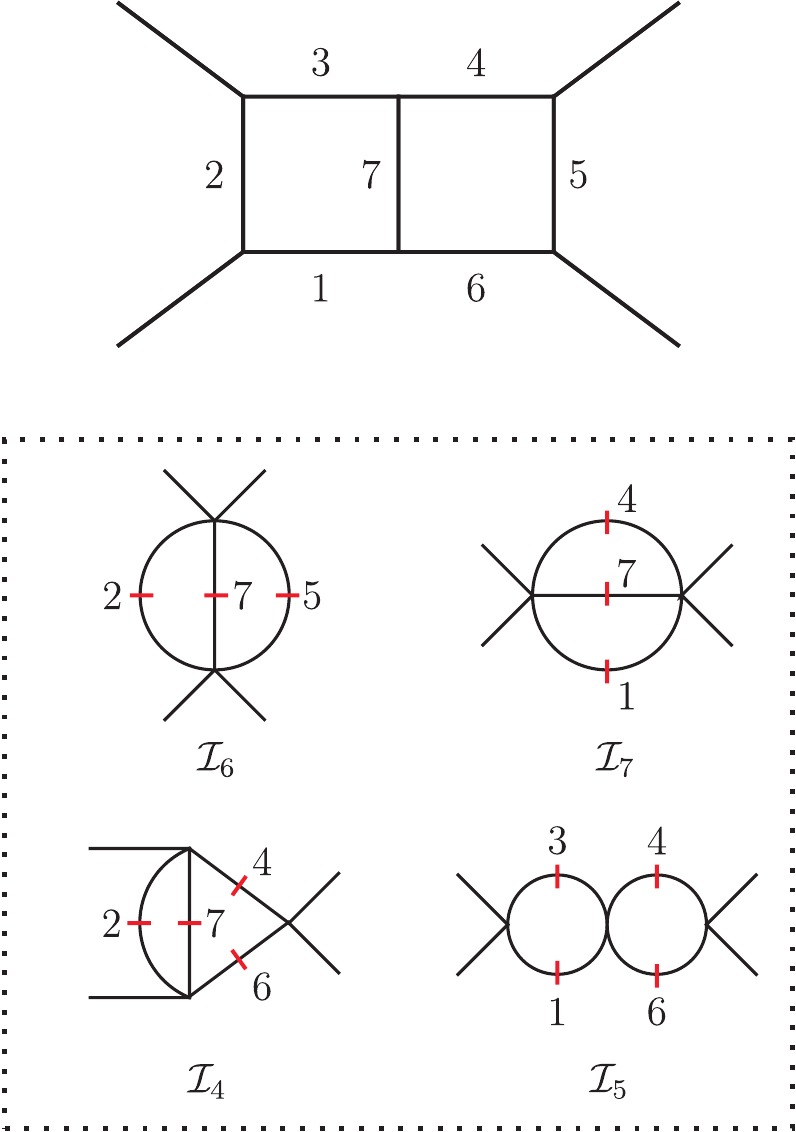}
\caption{The massless double-box diagram, along with our labelling
conventions for its internal lines, is shown on top. The lower
part shows the subset of the master integrals in Eq.~(\ref{eq:master_integrals})
with the property that their graphs cannot be obtained by
adding internal lines to the graph of some other master integral.
The corresponding cuts $\{2,5,7\}$, $\{1,4,7\}$, $\{2,4,6,7\}$
and $\{1,3,4,6\}$ are the cuts required for deriving complete
IBP relations for the double-box diagram.}
\label{fig:dbox_cut}
\end{figure}

To demonstrate the method, we consider the example of a double-box integral
with all legs and propagators massless, illustrated in Fig.~\ref{fig:dbox_cut}.
For this integral we have $k=7$, and the inverse propagators can be written as,
\begin{gather}
  D_1 =\ell_1^2,\quad D_2 =(\ell_1-p_1)^2,\quad D_3
  =(\ell_1-p_1-p_2)^2 \nonumber \\
D_4 =(\ell_2-p_3-p_4)^2,\quad D_5 =(\ell_2-p_4)^2,\quad D_6
  =\ell_2^2\nonumber \\
D_7=(\ell_1+\ell_2)^2 \,. \label{eq:3}
\end{gather}
As mentioned above Eq.~(\ref{eq:2}), the generic integral
of this topology will have numerator insertions which are
monomials in two distinct ISPs. The ISPs may be chosen as,
\begin{gather}
D_8=(\ell_1+p_4)^2/2,\quad D_9=(\ell_2+p_1)^2/2 \,.
\label{eq:5}
\end{gather}
Our aim is now to show how the IBP reductions of a generic
integral with the propagators in Eq.~(\ref{eq:3}) can be obtained.
We start by parametrizing the four-dimensional part of the loop momenta
in terms of van Neerven-Vermaseren's parametrization \cite{vanNeerven:1983vr},
\begin{align}
  \label{eq:6}
x_i&=\ell_1 \cdot p_i, \hspace{8mm} x_4 =\ell_1 \cdot \omega \nonumber \\
y_i&=\ell_2 \cdot p_i, \hspace{8mm} y_4 =\ell_2 \cdot \omega \,,
\end{align}
where $i=1,2,3$. Next, we introduce variables $\lambda_{11}, \lambda_{22}$
and $\lambda_{12}$ (cf.~Eq.~(\ref{lambda})), and integrate out $x_4$ and $y_4$.
We then re-parametrize the loop momenta as,
\vspace{-1mm}
\begin{gather}
  \label{eq:7}
  D_i=z_i, \quad i=1,\ldots,9 \,.
\end{gather}
The map $\{x_i, y_i, \lambda_{jk}\} \to z_i$
has a constant Jacobian and a polynomial inverse.

We will use the following notation for the integrals,
\vspace{-2mm}
\begin{gather}
  \label{eq:4}
  G[n_1,\ldots,n_9] \equiv \int \prod_{i=1}^{9} \d z_i \hspace{0.6mm}
  F(z)^{\frac{D-6}{2}} z_1^{n_1} \cdots z_9^{n_9} \,.
\end{gather}

Now, to find the IBP reductions of these integrals,
the first step is to find the set of master integrals. This is done by solving the
syzygy equation (\ref{syz}) without cuts using numerical external kinematics
(with rational or finite-field coefficients), then inserting all solutions into the right-hand side
of Eq.~(\ref{IBP}), and finally row reducing.
In the case at hand, we find the following set of master integrals
(after modding out by symmetries)
\begin{equation}
\begin{array}{rl}
\mathcal{I}_1 \hspace{-0.5mm} &\equiv \hspace{0.5mm} G[-1,-1,-1,-1,-1,-1,-1,0,0]\\
\mathcal{I}_2 \hspace{-0.5mm} &\equiv \hspace{0.5mm} G[-1,-1,-1,-1,-1,-1,-1,1,0]\\
\mathcal{I}_3 \hspace{-0.5mm} &\equiv \hspace{0.5mm} G[0, -1, -1, 0, -1, -1, -1, 0, 0]\\
\mathcal{I}_4 \hspace{-0.5mm} &\equiv \hspace{0.5mm} G[0, -1, 0, -1, 0, -1, -1, 0, 0] \\
\mathcal{I}_5 \hspace{-0.5mm} &\equiv \hspace{0.5mm} G[-1, 0, -1, -1, 0, -1, 0, 0, 0] \\
\mathcal{I}_6 \hspace{-0.5mm} &\equiv \hspace{0.5mm} G[0, -1, 0, 0, -1, 0, -1, 0, 0]\\
\mathcal{I}_7 \hspace{-0.5mm} &\equiv \hspace{0.5mm} G[-1, 0, 0, -1, 0, 0, -1, 0, 0] \\
\mathcal{I}_8 \hspace{-0.5mm} &\equiv \hspace{0.5mm} G[-1, -1, -1, 0, -1, 0, -1, 0, 0] \,.
\end{array}
\label{eq:master_integrals}
\end{equation}
Having obtained the set of master integrals, we proceed to find
the IBP reductions analytically. This step is done most efficiently
by determining the reductions on a set of cuts and then combining
the results to find the complete reductions. To decide on the
minimal set of cuts required, we consider the subset of master integrals
with the property that their graphs cannot be obtained by
adding internal lines to the graph of some other master integral.
In the case at hand, this subset is $\{ \mathcal{I}_4, \mathcal{I}_5,
\mathcal{I}_6, \mathcal{I}_7\}$, shown in Fig.~\ref{fig:dbox_cut}.
Hence, we only need to consider the four cuts $\{2,5,7\}$, $\{1,4,7\}$,
$\{2,4,6,7\}$ and $\{1,3,4,6\}$ to find the complete IBP reductions.

To see how to find the IBP reductions on a given cut,
let us consider the three-fold cut $\mathcal{S}_\mathrm{cut}=\{2,5,7\}$.
Here, $\mathcal{S}_\mathrm{uncut}=\{1,3,4,6\}$ and $\mathcal{S}_\mathrm{ISP}=\{8,9\}$. The
kernel $F$ on the cut is
polynomial in $z_1$, $z_3$, $z_4$, $z_6$, $z_8$ and $z_9$. The syzygy equation
(\ref{syz}) reads,
\begin{align}
bF -\sum_{i\in\{ 1,3,4,6\}} b_i z_i \frac{\partial F}{\partial z_i} \hspace{1mm}
+ \sum_{j\in\{ 8,9\}} a_j \frac{\partial F}{\partial z_j} =0 \,,
 \label{dbox_sunset1_syz}
\end{align}
where $b, b_i, a_j$ are to be solved for as polynomials in $z_k$.
The solutions of Eq.~(\ref{dbox_sunset1_syz}) can
be found via algebraic geometry software such as Macaulay2 or Singular
in seconds (with analytic coefficients). Now, given
a solution $(b, b_i, a_j)$, any multiple
$(q b, q b_i, q a_j)$, with $q$ a polynomial,
is also a solution. To capture the IBP reductions of
all possi\-ble numerator insertions, we thus consider all syzygies
$(b, b_i, a_j)$ multiplied by appropriate
monomials in the ISPs, $q = \prod_{i \in \{ 1,3,4,6,8,9 \}} D_i^{a_i}$.
Inserting all such solutions into the right-hand side
of Eq.~(\ref{IBP}) produces the full set of
IBP relations, without doubled
propagators, on the cut $\{2,5,7\}$ (i.e., up to integrals that vanish on this cut).

As an example, consider the tensor integral $T\equiv G[-1,-1,-1,-1,-1,-1,-1,0,2]$.
On the four cuts this integral reduces to, respectively,
\begin{align}
T\big|_{\{2,5,7\}}   &= \hspace{-2mm} \sum_{j\in \{ 1,2,3,6,8 \}} \hspace{-3mm}
c_j \mathcal{I}_j \,, \hspace{2mm} T\big|_{\{1,4,7\}}    = \hspace{-2mm}
\sum_{j\in \{ 1,2,7 \}} \hspace{-2mm} c_j \mathcal{I}_j \,,
\label{eq:IBP_on_cut_formal_1}\\
T\big|_{\{2,4,6,7\}} &= \hspace{-2mm} \sum_{j\in \{ 1,2,4 \}} \hspace{-2mm}
c_j \mathcal{I}_j \,, \hspace{2mm} T\big|_{\{1,3,4,6\}}  = \hspace{-2mm}
\sum_{j\in \{ 1,2,5 \}} \hspace{-2mm} c_j \mathcal{I}_j \,,
\label{eq:IBP_on_cut_formal_2}
\end{align}
where, using $\chi \equiv t/s$, the coefficients are found to be,
\begin{align}
c_1 &= \frac{(D-4) s^2\chi}{8 (D-3)}\,, \hspace{2mm}
c_2 = -\frac{(3 D-2 \chi -12)s}{4 (D-3)} \\
c_3 &= \frac{(4-D) (9 \chi +7)}{4 (D-3)} \,, \hspace{2mm}
c_4 = \frac{(10 -3D) (2 \chi -13)}{8 (D-4)s} \\
c_5 &= \frac{2 D (\chi +1){-}8 \chi{-}7}{2 (D-4)s} \,, \hspace{2mm}
c_6 = \frac{9 (3D{-}10) (3D{-}8)}{4 (D-4)^2 s^2\chi} \\
c_7 &= \frac{(3 D-10) (3 D-8) (2 \chi +1)}{2 (D-4)^2 (D-3) s^2} \,, \hspace{2mm}
c_8 = 2 \,.
\end{align}
The integrals absent from the right-hand sides of
Eqs.~(\ref{eq:IBP_on_cut_formal_1})--(\ref{eq:IBP_on_cut_formal_2})
vanish on the respective cuts. Combining these results,
we get the \emph{complete} IBP reduction of the tensor integral,
\begin{equation}
T = \sum_{j=1}^8 c_j \mathcal{I}_j \,.
\label{eq:complete_IBP_of_tensor}
\end{equation}

We have implemented the algorithm as a program, powered by Mathematica,
Macaulay2~\cite{M2}, Singular~\cite{DGPS} and Fermat~\cite{Fermat}. It analytically
reduces all integrals with numerator
rank $\leq 4$, to the eight master integrals in
Eq.~(\ref{eq:master_integrals}) in about $39$ seconds in the fully
massless case, and to $19$ master integrals in about $211$ seconds in the
one-massive-particle case (on a laptop with 2.5 GHz Intel Core i7 and 16 GB RAM).

One important feature of the approach developed in this paper
is the use of the $z_i$-variables in Eq.~(\ref{eq:2}) which
which ultimately generate the simple syzygy equations (\ref{eq:syzygy_1})--(\ref{eq:syzygy_2}).
Another feature is the use of unitarity cuts: they eliminate
variables in the syzygy equations so that these can be solved
more efficiently. There are several directions
for future research. Of particular interest are extensions to
higher multiplicity, several external and internal masses,
non-planar diagrams, and higher loops.

{\bf Acknowledgments}

We thank S.~Badger, H.~Frellesvig, E.~Gardi, A.~Georgoudis, A.~Huss, H.~Ita,
D.~Kosower, A.~von~Manteuffel, M.~Martins, C.~Papadopoulos and R.~Schabinger for useful discussions.
The research leading to these results has received
funding from the European Union Seventh Framework
Programme (FP7/2007-2013) under grant agreement no.~627521, and Swiss
National Science Foundation (Ambizione~PZ00P2\_161341).

\bibliography{IBPs_from_syzygies}

\end{document}